\documentclass[twocolumn,showpacs,preprintnumbers,amsmath,amssymb]{revtex4}
\usepackage{graphicx}% Include figure files
\usepackage{dcolumn}% Align table columns on decimal point
\usepackage{bm}% bold math

\begin{document}

%\preprint{APS/123-QED}

\title{Adaptive network models of collective decision making in swarming systems}
%\title{Collective decision making with adaptive network models}
%\title{Adaptive network description of swarming systems with multiple heading directions}
%\title{Multi-dimensional collective decision making with adaptive network models}
% Force line breaks with \\

\author{Li Chen}
\email{chenli@pks.mpg.de}
\affiliation{Max Planck Institute for the Physics of Complex Systems, 01187 Dresden, Germay}
%\affiliation{Project group 4,Robert  Koch Institute, 13353 Berlin, Germany}
\author{Cristi\'an Huepe}
%\email{chenli@pks.mpg.de}
\affiliation{CHuepe Labs, 922 W 18th Place, Chicago, Illinois 60608, USA}
\affiliation{Northwestern Institute on Complex Systems, Northwestern University, Evanston, Illinois 60208, USA}
\author{Thilo Gross}
%\email{chenli@pks.mpg.de}
\affiliation{Department of Engineering Mathematics, Merchant Venturers Building, University of Bristol, Woodland Road, Clifton, Bristol BS8 ITR, United Kingdom}

%\date{\today}
% It is always \today, today, but any date may be explicitly specified

\begin{abstract}
We consider a class of adaptive network models where links can only be created or deleted between nodes in different states. These models provide an approximate description of a set of systems where nodes represent agents moving in physical or abstract space, the state of each node represents the agent's heading direction, and links indicate mutual awareness. We show analytically that the adaptive network description captures the phase transition to collective motion in swarming systems and that the properties of this transition are determined by the number of states (discrete heading directions) that can be accessed by each agent.
\end{abstract}

%PACS: Physics and Astronomy Classification Scheme.
%   05.90.+m: Other topics in statistical physics
%   89.75.Hc: Networks and genealogical trees
%   87.23.Cc: Population dynamics and ecological pattern formation
%\keywords{Suggested keywords}
%   Use showkeys class option if keyword display desired

\maketitle

%===============================================================
\section{Introduction}
%===============================================================

% Introduce Adaptive Networks
Adaptive networks define a versatile class of models that have been recently applied to a wide variety of systems \cite{GrossReview,GrossSayama}. They combine processes that change the structure of a network, such as growth or rewiring, with dynamics taking place on the network. This results in a feedback between topology and dynamics that can lead to different forms of self-organization. Following the pioneering work of Bornholdt and Rohlf \cite{BornholdtAndRohlf} adaptive networks have been applied to a wide range of systems, including neural networks \cite{Levina,Meisel}, mobile sensor networks \cite{DiBernardo,Leonard}, epidemics \cite{Gross,ShawSchwartz}, and the evolution of cooperation \cite{Traulsen,Do}, among many others \cite{ANwiki}.

% Introduce adaptive voter model
In the study of adaptive networks, a special role is played by opinion formation models and, in particular, by the adaptive voter model and its variants \cite{Holme,Federico,Gil,SanMiguel,Kimura,Gesa1,Gesa2,Demirel}. The adaptive voter model describes the process through which a population of agents forms an opinion. A group of nodes representing agents are connected by links that describe social interactions. Each node is associated to a variable that can take values representing all possible opinions. At every iteration, the network is updated by propagating these values along the links (social adjustment) and by rewiring links (social segregation).
% Introduce homophily & heterophily
One typically considers nodes that rewire their connections to surround themselves by like-minded agents that hold the same opinion. This common type of social dynamics is called \emph{homophily}. Its opposite \emph{heterophily}, where agents seek connections to different-minded agents \cite{Kimura}, has received much less attention.

% Adaptive voter & collective motion
Extensions of the adaptive voter model have been recently proposed to describe collective motion in groups of animals \cite{Cristian,Couzin2011}, a basic social phenomenon that occurs in a broad variety of species. Examples include insect swarms, fish schools, bird flocks, herds of quadrupeds \cite{Couzin}, and even crowds of people \cite{Helbing}. 
Here, we will refer to all these, generically, as {\it swarming systems}.
The process through which such systems self-organize into coordinated collective motion is still poorly understood. There has been much debate, for example, regarding the nature of the swarming transition that marks the onset of collective motion \cite{VicsekPR}. 

% Adaptive network description of swarms: basics and literature
Most theoretical studies investigate swarming by either analyzing detailed agent-based models  \cite{Vicsek1} or representing the swarm as a continuous medium \cite{TT1,TT2}.
Adaptive network models provide an alternative route: the swarm is represented as an adaptive network by an approximation that captures the agents' headings and interactions but neglects their trajectories in space.
In such models, each agent is represented by a node, its heading direction is treated as an internal state, and mutual awareness between two agents is represented by a link.
But, there is no explicit representation of space, i.e.~no variable keeps track of each agent's position in space. 

Network swarming models are reminiscent of the standard adaptive voter model if we view the heading directions as the different potential opinions in an opinion formation process.
A notable difference, however, is that in the swarming case having different opinions (i.e. moving in different directions) increases, both, the probabilities of creating and of destroying links between agents
 (Fig.~\ref{fig:1}). 
The adaptive swarming models thus constitute a third class of opinion formation systems comprising aspects of both homophily and heterophily. We refer to such systems as the {\it swarming systems class} of adaptive network models.
This type of approach was originally proposed \cite{Cristian} to model experiments on the collective motion of groups of locusts marching on a ring-shaped arena \cite{Buhl}. A slightly extended version of this model was later used to predict the outcome of decision-making experiments with fish \cite{Couzin2011}.

% Special homophily and heterophily combination in collective motion adaptive network case & corresponding social network case
We note that the swarming systems class of adaptive network models may also be relevant for other applications that consider motion in abstract (rather than physical) space. For instance, if translated into a social context where different heading directions correspond to different opinions, it describes individuals that create or destroy social connections mainly with those who have a different opinion. While this is not the most common social dynamics, it may describe situations where original opinions are strongly valued and attract new social interactions but also create tensions within established interactions, leading to dissolution.

% What we do here
The previously proposed adaptive network models for swarming systems considered only cases where each agent was restricted to choose between two heading directions, corresponding to clock-wise or counter-clockwise motion on the circular arena \cite{Cristian} or to swimming towards one of two targets \cite{Couzin2011} in the collective decision-making experiment. These investigations thus focused on situations where the internal opinion state was a binary variable, akin to the adaptive voter model. While several multi-state extensions to the voter model have been explored \cite{Deffuant,Holme,Gesa2,Durret}, the present paper is the first to analyze a similar extension for the swarming system class of adaptive network models. 
This extension is intuitive, as real swarms typically move in two or three-dimensional space, where the heading direction can be discretized into more than two node states. 

% What we do in this paper
In this paper we show that the swarming system class of adaptive network models displays a symmetry-breaking ordering transition that can be likened to collective motion. This transition can be either continuous or discontinuous, depending on the number of accessible states (e.g.~the dimensionality of the embedding space).

% Paper organization
The paper is organized as follows. Section \ref{Sec:Model} introduces the swarming system class of adaptive network models. Section \ref{Sec:MeanField} analyzes its mean field approximation. Section \ref{Sec:AdaptiveNetworkSol} computes its analytical and numerical solutions. Section \ref{Sec:Swarm} compares our adaptive network results with the standard swarming transition to collective motion. Finally, Section \ref{Sec:Conclusions} presents our conclusions.

%===============================================================
\section{Adaptive network system}
\label{Sec:Model}
%===============================================================

%------------------------------------------------------------------------------------------ Figure 1 ------------------------
\begin{figure}[tp]
\includegraphics[scale=0.4]{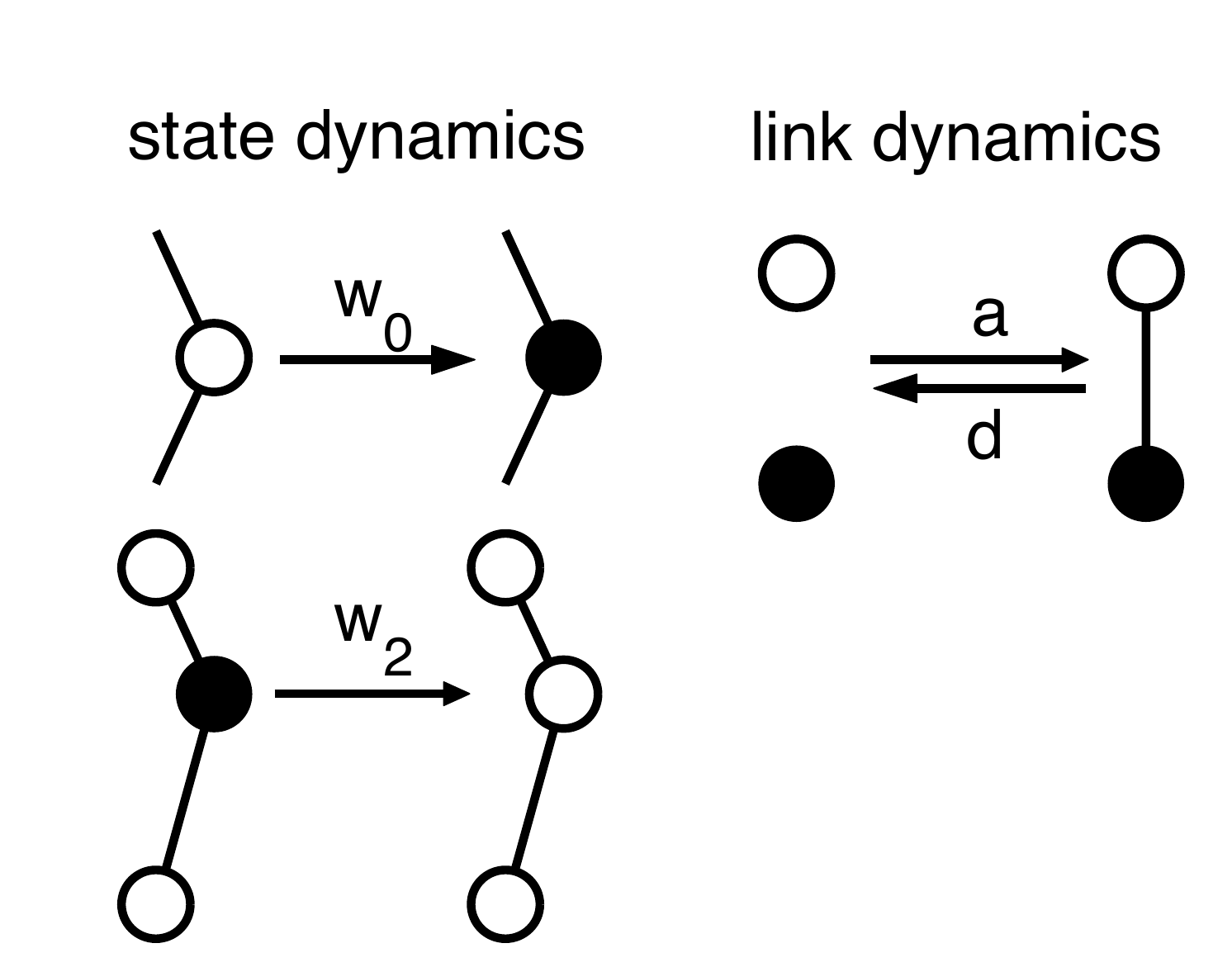}
\caption{Model illustration.
The diagram presents nodes (circles) displaying two different opinions (black/white) out of $M$ possible choices. 
The state dynamics (left column) consists of spontaneous fluctuations of individual nodes (top) and of three-body processes (bottom), with rates $w_0$ and $w_2$, respectively. 
The link dynamics (right column) consists of the creation and deletion of links only between nodes in different states, with rates $a$ and $d$, respectively.
These dynamics take place irrespective of any additional links, which may be present but are not shown in this figure, or of the total number of links connected to the node in the $w_0$ process.
}
\label{fig:1}
\end{figure}
%----------------------------------------------------------------------------------------------------------------------------- 

We consider a system of $N$ nodes, representing agents, connected by links representing mutual awareness. Each agent has an internal variable that encodes its opinion state (or, equivalently, its heading direction) as one of $M$ potential discrete states. For convenience, we denote the set of all possible opinion states by $\Omega\!=\!\{\!1,2,..., M\!\}$ and the complement of a given state $X$ with respect to $\Omega$ by $\overline{\Omega}_{\{\!X\!\}}=\Omega\!\setminus\!\{\!X\!\}$.
The initial states of the agents are drawn from $\Omega$ with equal probability. 
The network is initialized as an Erd\H{o}s-R\'enyi random graph with inital mean degree $\langle k\rangle=3$. The network then evolves in time as follows (see Fig.~\ref{fig:1}):

\emph{State dynamics} --- The state of each node is updated according to one of the following two processes. 
(\emph{i}) Every node changes its state spontaneously at a net rate of $w_0$ changes per node,
picking one of the $M-1$ other states in $\overline{\Omega}_{\{\!X\!\}}$ with equal probability. 
({\emph {ii}}) In every triplet of nodes $Y\!-X\!-Y$, where two nodes on the same state $Y$ are linked to a single node on a different state $X$, the central node switches its state to $Y$ with a probability that amounts to a net rate of $w_2$ transitions per triplet \cite{footnote1}.

\emph{Link dynamics} --- Links are established or removed only between pairs of nodes that are in different states, with probabilities that amount to a net creation and deletion rates of $a$ (per pair) and $d$ (per link), respectively.

All numerical  network simulations were carried our using an event-driven (Gillespie) algorithm that yields a very good approximation of the continuous-time dynamics at the link level \cite{ZschalerBioinformatics}.  
%
%-------------------------------------------------------------------------------------new Figure 2 ------------------------
\begin{figure}[htp]
\includegraphics[scale=0.35]{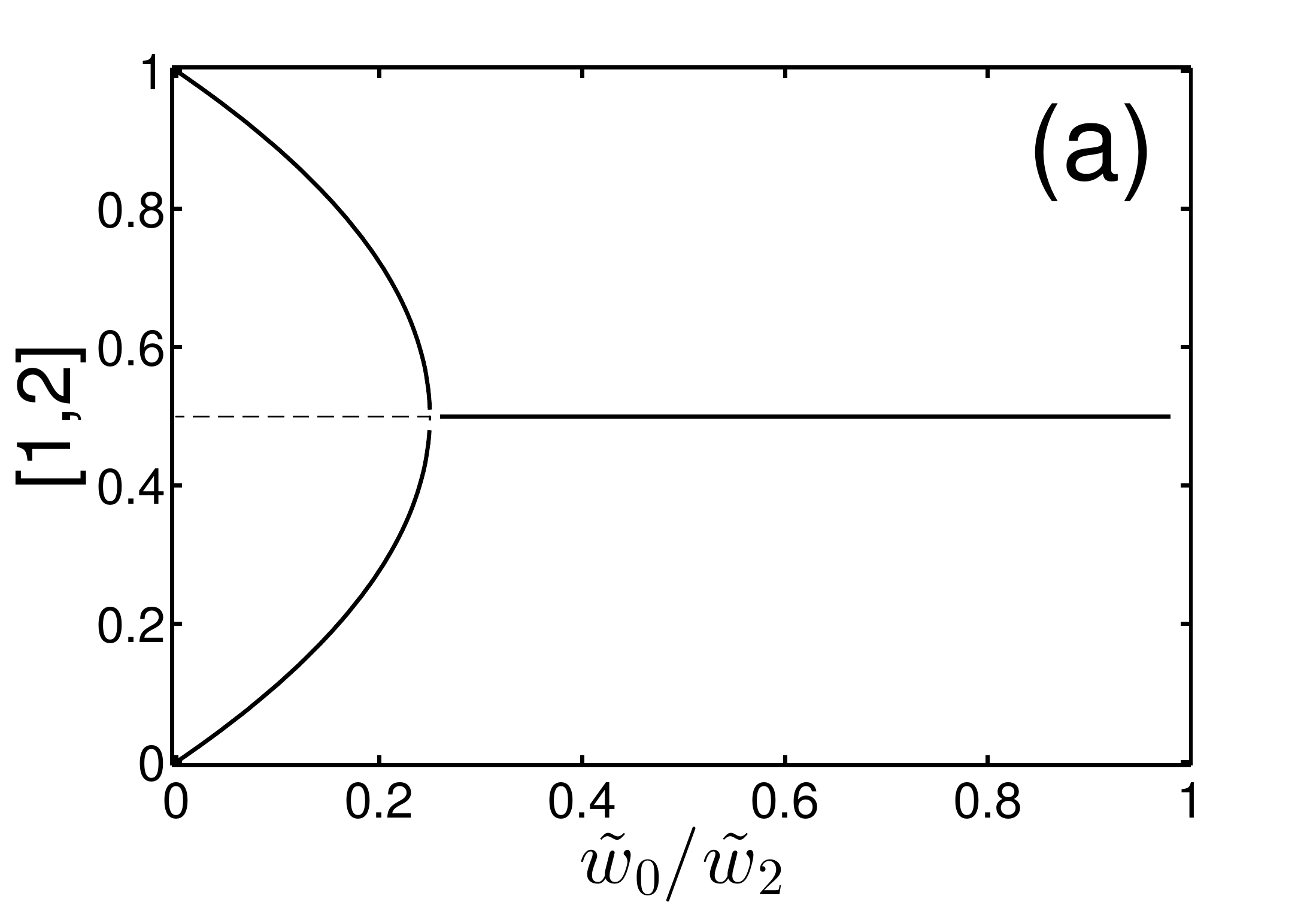}
\includegraphics[scale=0.35]{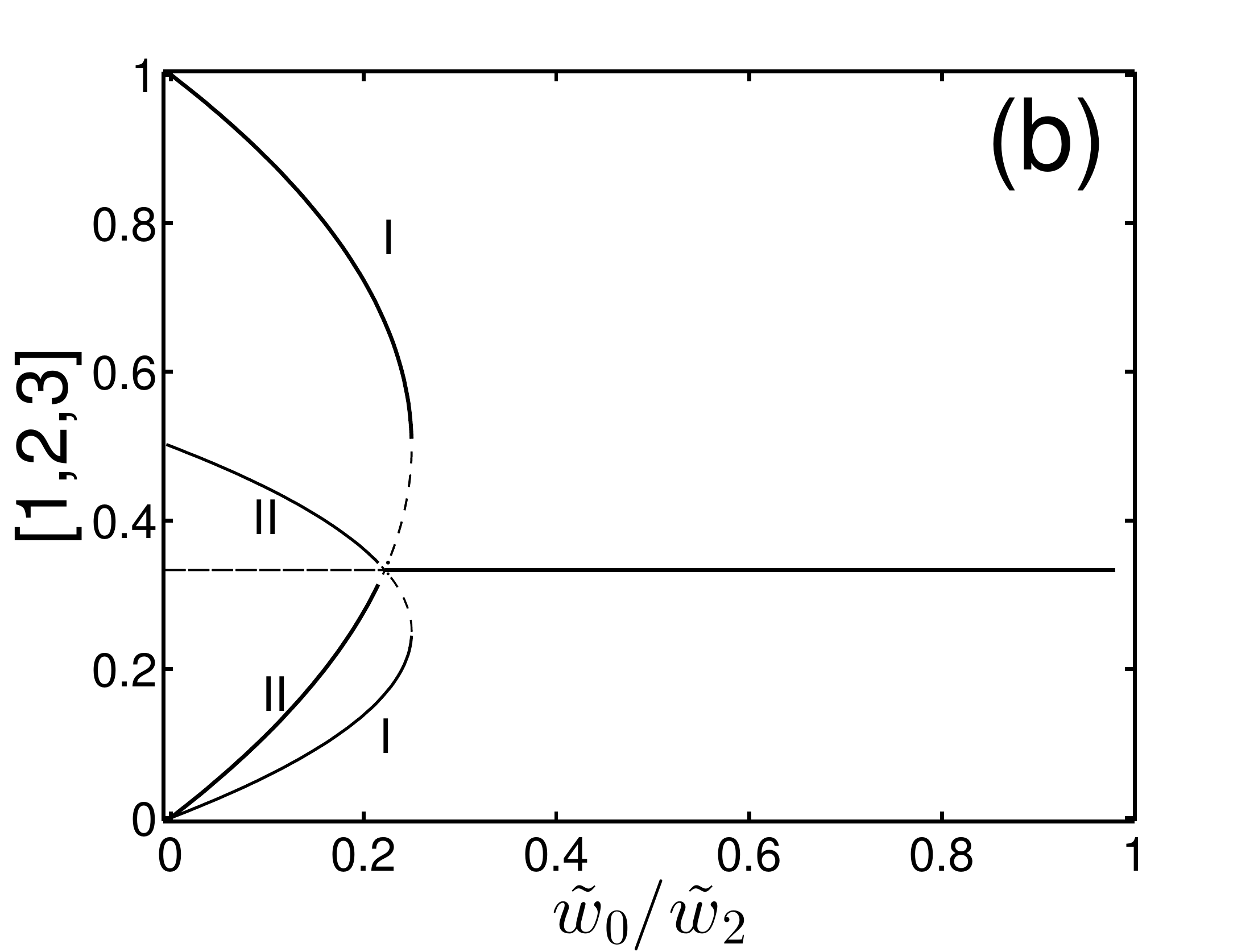}
\caption{ 
Bifurcation diagram of the mean field approximation of the density of agents in each state as a function of normalized noise $\tilde{w}_0/\tilde{w}_2$ in the $M=2$ (a) and $M=3$ (b) cases. 
The curves represent stable (solid) and unstable (dashed) branches of the steady state solutions.  
The $M=2$ case (a) undergoes a continuous transition in the form of a supercritical pitchfork bifurcation. 
The $M=3$ case (b), presents two sets of stable solutions: 
one set (I) appears through a discontinuous transition and corresponds to a single majority opinion and two minority opinions with the same number of agents,
the other set (II) results from a continuous transition and corresponds to two majority opinions with equal number of agents and a single minority opinion.
}
\label{fig:2}
\end{figure}
%-----------------------------------------------------------------------------------------------------------------------------

%===============================================================
\section{Mean field solution}
\label{Sec:MeanField}
%===============================================================

Before carrying out an adaptive network analysis, it is instructive to gain some intuition by considering a mean field approximation. 
This approximation is equivalent to neglecting the link dynamics and assuming that the density of links connecting nodes in given states is proportional to the product of the densities of these states. 
While crude, it leads to a picture that is qualitatively similar to the adaptive network results described in Section \ref{Sec:AdaptiveNetworkSol}.

For simplicity, we denote by $x$ the density of any given state and by $y_i$ the density of all other $M-1$ states. The mean field approximation then leads to
\begin{equation}
\label{eq:MeanFieldSystem}
\frac{{\rm d} x}{{\rm d} t} = \frac{w_0}{M-1} \left(\sum_{i=1}^{M-1} y_i\right) - w_0 x + 
						w_2 \langle k \rangle^2 \sum_{i=1}^{M-1} \left( x^2 y_i - y_i^2 x \right),
\end{equation}
where $\langle k \rangle$ is the mean degree, i.e.~the mean number of links per network node.
 
In the equation the first two terms describe the gain and loss of nodes in state $x$ due to spontaneous switching, respectively, and the last term captures the gains and losses resulting from the triplet process. The conservation of the total node density implies that $x+\sum_i y_i =1$.
Basic intuition and preliminary numerical simulations suggest that the system will either converge towards a disordered (mixed) solution where all node states occur with the same probability or to an ordered solution where a preferred direction emerges and its corresponding state is overexpressed in the population, while the other states remain at an equal, lower density. We can thus make analytical progress by assuming $y_1=y_2=\ldots=:y$. This leads to the simplified system
\begin{equation}
\label{eq:MeanField}
\frac{{\rm d} x}{{\rm d} t} = \tilde{w}_0 \left( y - x \right) + \tilde{w}_2 \left( M-1 \right) \left( x^2 y - y^2 x \right),
\end{equation}
where we defined $\tilde{w}_0=w_0$ and $\tilde{w}_2=w_2 \langle k \rangle^2$, to simplify the expression.
We now compute the steady state solutions of this system by setting the left-hand side of Eq.~(\ref{eq:MeanField}) to zero. Factorizing $y-x$ we obtain
\begin{equation}
0=(y-x) [ \tilde{w}_0-\tilde{w}_2(M-1)xy ].
\end{equation}
From this equation it is apparent that we get a symmetric solution $x=y$ and asymmetric solutions that satisfy 
\begin{equation}
xy=\frac{\tilde{w}_0}{\tilde{w}_2(M-1)}.
\end{equation}
Using the normalization condition $(M\!-\!1)y+x=1$, we find that the symmetric solution is given by $x = 1/M$, and the asymmetric pair by 
\begin{equation}
\label{eq:parabolic}
x = \frac{1}{2} \pm \sqrt{ \frac{1}{4} - \frac{\tilde{w}_0}{\tilde{w}_2} },
\end{equation}
which is independent of the number of states $M$. The constant symmetric solution represents a disordered state where all heading directions are equally probable. The parabolic asymmetric solutions in Eq.~(\ref{eq:parabolic}), ordered cases with preferred heading directions.

As the noise level is increased, the system undergoes a transition from the ordered to the disordered state (Fig.~\ref{fig:2}). 
Even before carrying out a linear stability analysis \cite{Kuznetsov}, direct visual inspection reveals the bifurcation points at which the stability of these steady state branches changes: 
Bifurcations occur both at the tips of the parabolas and at the intersection point of the different solutions. 

The tip of the parabola corresponds to the point $\tilde{w}_0/\tilde{w}_2=1/4$, where the stable and unstable solution branches meet through a saddle-node bifurcation.
The intersection of the two solutions occurs at $\tilde{w}_0/\tilde{w}_2 = (1-1/M)/M$ where a degenerate transcritical bifurcation takes place.  

In the context of the full system, the bifurcation points correspond to phase transitions. 
For any $M > 2$, the destabilization of the mixed state occurs through a subcritical bifurcation, corresponding to a discontinuous transition.
Only in the $M=2$ case the two bifurcation points coincide at $\tilde{w}_0/\tilde{w}_2 = 1/4$ and become a supercritical pitchfork bifurcation, corresponding to a continuous transition. 

A detailed stability analysis \cite{XPPAUT} of Eq.~(\ref{eq:MeanField}) verifies the results above and shows an additional set of stable solution branches in Fig.~\ref{fig:2}(b) (labeled by II). In these branches two majority opinions are represented in an equal number of nodes and while a single minority opinion is held by a smaller number of nodes.
However, in the next section we show that the stability of these branches is lost when a more accurate approximation is used. 
This suggests that the stability of the 2-majority/1-minority branches is a spurious result of the mean field approximation, which appears due to an excessive reduction of the dimensionality of the state space. 
In the full system these branches must thus be unstable to certain perturbations that involve a dynamical redistribution of links, which is not captured by the mean field. 
%
% XXXXXXXXXXXXXXXXXXXXXXXXXXXXXXXXXXXXXXXXXXXXXXXXXXXXXXXXXXXXXXXXXXXXXXXXXXXXXX

%===============================================================
\section{Adaptive Network Solution with Pair-level Closure}
%Link-level Solution of Adaptive Networks}
\label{Sec:AdaptiveNetworkSol}
%===============================================================

We now derive a system of equations that takes the dynamics of link densities into account, using a moment expansion \cite{Demirel2}.
The basic idea of this expansion is to write differential equations that capture the density of small subgraphs. These densities are also called network moments. Each subgraph can be classified by its order, which is equal to the number of links it contains. For example, if we have three distinct states $X\!,\!Y\!,\!Z\!\in\!\Omega$, the density of nodes in the $X$ state, denoted by $[X]$,  is a zeroth-order moment; the per-capita density of $X\!-\!Y$ linked pairs $[XY]$, a first-order moment; and the $X\!-\!Y\!-\!Z$ triplet density $[XYZ]$, a second-order moment.
With these definitions, the dynamics of the zeroth and first order moments are captured by
\begin{widetext}
\small
\begin{equation}
\label{Eq:AN1}
\frac{\rm d}{\rm dt}[X]=\frac{w_{0}}{M\!\!-\!\!1}\Big \{\!\!\!\sum_{Y\!\in\overline{\Omega}_{\{\!X\!\}}}\!\!\!\![Y]-\!(M\!\!-\!\!1)[X]\!\Big \}\!+\!w_{2}\!\!\sum_{Y\!\in\overline{\Omega}_{\{\!X\!\}}}\!\!\!\! \Big \{ [X\!Y\!X]\!-\![Y\!X\!Y] \Big \},
\end{equation}
\begin{equation}
\label{Eq:AN2}
\begin{split}
\frac{\rm d}{\rm dt}[X\!X]
&=\frac{w_{0}}{M\!\!-\!\!1}\Big \{\!\!\sum_{Y\!\in\overline{\Omega}_{\{\!X\!\}}}\!\!\!\![X\!Y]\!-\!2(M\!\!-\!\!1)[X\!X]\Big \}\!
\!+\!w_{2}\!\!\!\sum_{Y\!\in\overline{\Omega}_{\{X\}}}\!\!\!\!\Big \{2[X\!Y\!X]\!+\!3[^XY_X^X]\!-\![^XX_Y^Y]\Big \},  \\
\end{split}
\end{equation}
\begin{equation}
\label{Eq:AN3}
\begin{split}
\frac{\rm d}{\rm dt}[X\!X']
&=\frac{w_{0}}{M\!\!-\!\!1}\Big \{2([X\!X]\!+\![X'\!X'])\!+\!\!\!\!\sum_{Y\!\in\overline{\Omega}_{\!\{\!X\!,\!X\!'\!\}\!}}\!\!\!\!([XY]\!+\![X'Y])\!-\!2(M\!\!-\!\!1)[X\!X']\Big \}\!+\!w_{2} \Big \{-2[X\!X\!'\!X]-2[X'\!X\!X']+[^X\!X_{X'}^{X'}] \\
&+\![^{X\!'}\!{X\!'}_{X}^{X}]\!-\!3[^{X\!'}\!X_{X'}^{X'}]\!-\!3[^X{X'}_{X}^{X}]\!+\!\!\!\!\sum_{Y\!\in\overline{\Omega}_{\{\!X\!,\!X\!'\!\}}}\!\!\!\!\big([^{X'}\!Y_{X}^{X}]\!+\![^X\!Y_{X'}^{X'}]\!-\![^X\!{X'}_{Y}^{Y}]\!-\![^{X'}\!X_{Y}^{Y}]\big)\Big \}\!+\!a[X][X']\!-\!d[X\!X'],
\end{split}
\end{equation}
\normalsize
\end{widetext}
where $X,X'\!\in\!\Omega$ (with $X\!\neq\!X'$), and $[^X\!Y_{Z}^{W}]$ denotes the density of motifs with a central node in state $Y$ connected to three other nodes in states $X$, $W$, and $Z$.
In all these equations, the first right-hand side term corresponds to noise-driven state dynamics and the second, to three-body interactions. The remaining terms in Eq.~(\ref{Eq:AN3}) result from the link creation and deletion processes. These expressions summarize a larger system of equations, with (\ref{Eq:AN1}), (\ref{Eq:AN2}) and (\ref{Eq:AN3}) representing $M$, $M$, and $M(M-1)/2$ equivalent equations, respectively.
Note that each equation describing the dynamics at a given order involves higher order terms. We thus need to close the system through a moment closure approximation. We use a pair-level closure \cite{Keeling,Gross} of the form
%
%\begin{eqnarray}
\begin{equation}
\label{Eq:ANtri}
[X\!Y\!Z]=\frac{h([X\!Y])h([Y\!Z])}{h([X\!Y\!Z])}\!\frac{[X\!Y][Y\!Z]}{[Y]}, \nonumber 
\end{equation}
\begin{equation}
\label{Eq:ANqua}
[^X\!Y\!^Z_W]=\frac{h([X\!Y])h([Y\!Z])h([Y\!W])}{h([^X\!Y\!^Z_W])}\!\frac{[X\!Y][Y\!Z][Y\!W]}{[Y]^2}, \nonumber
%\end{eqnarray}
\end{equation}
where $h([X\!Y])=\!1\!+\!\delta_{XY}$, $h([X\!Y\!Z])=\!1\!+\!\delta_{XZ}$ and
$h([^XY^Z_W])=\!1\!+\!\delta_{XZ}\!+\!\delta_{XW}\!+\!\delta_{ZW}\!+\!\delta_{XZ}\delta_{ZW}\!+\!\delta_{XW}\delta_{ZW}$, with $\delta$ the Kronecker delta.
%

%------------------------------------------------------------------------------------------ Figure 3 ------------------------
\begin{figure}[htp]
\includegraphics[scale=0.225]{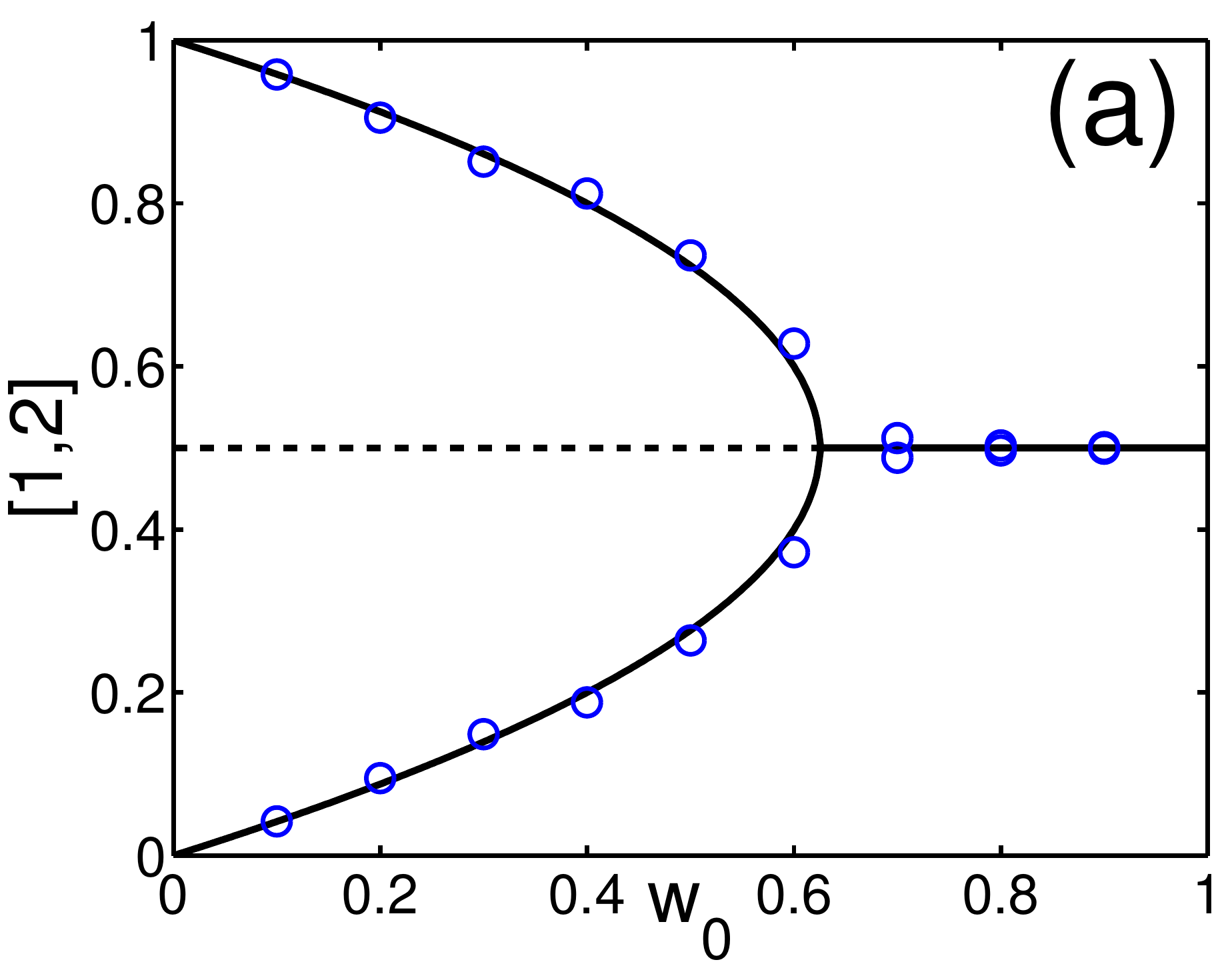}
\includegraphics[scale=0.225]{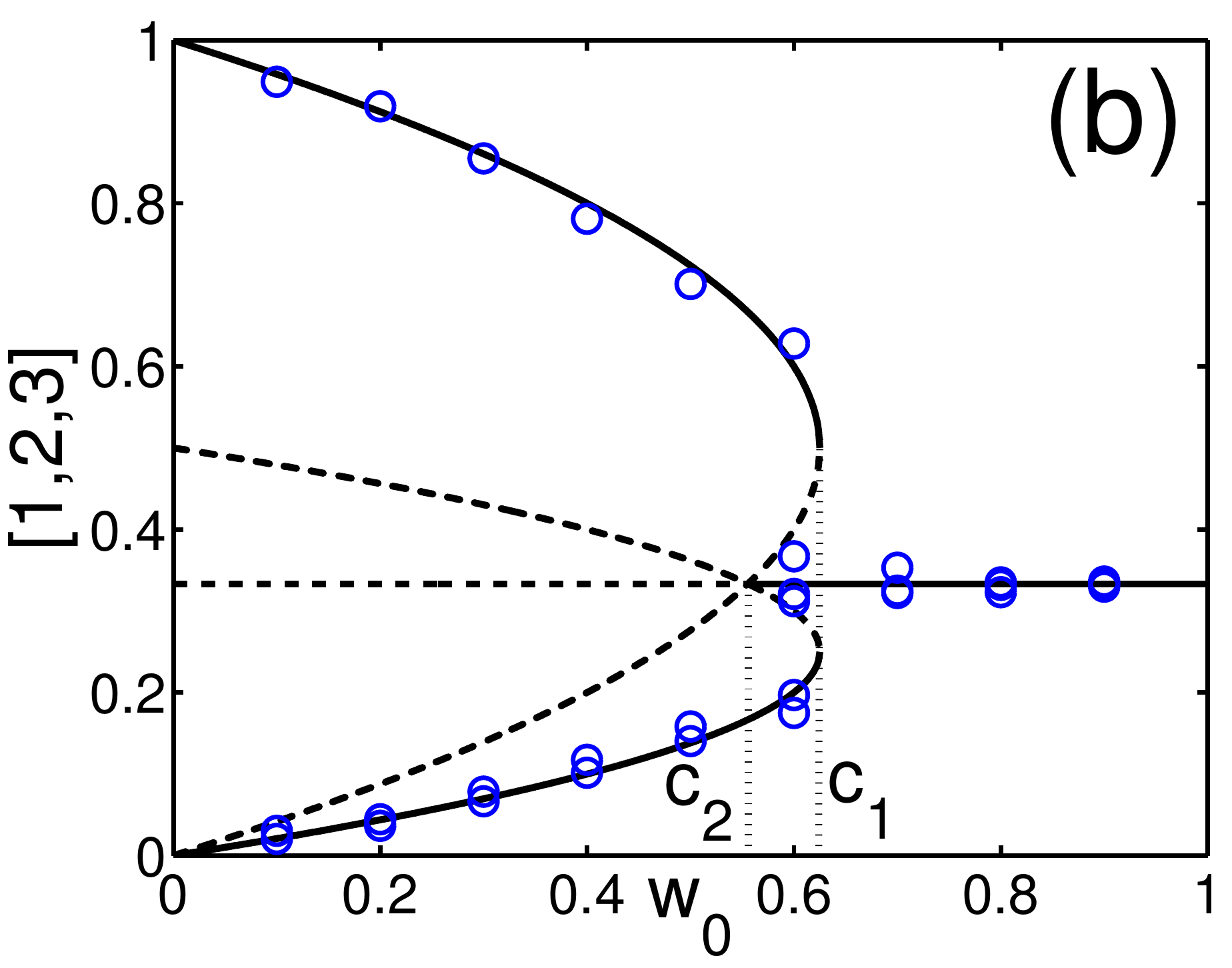}
\includegraphics[scale=0.225]{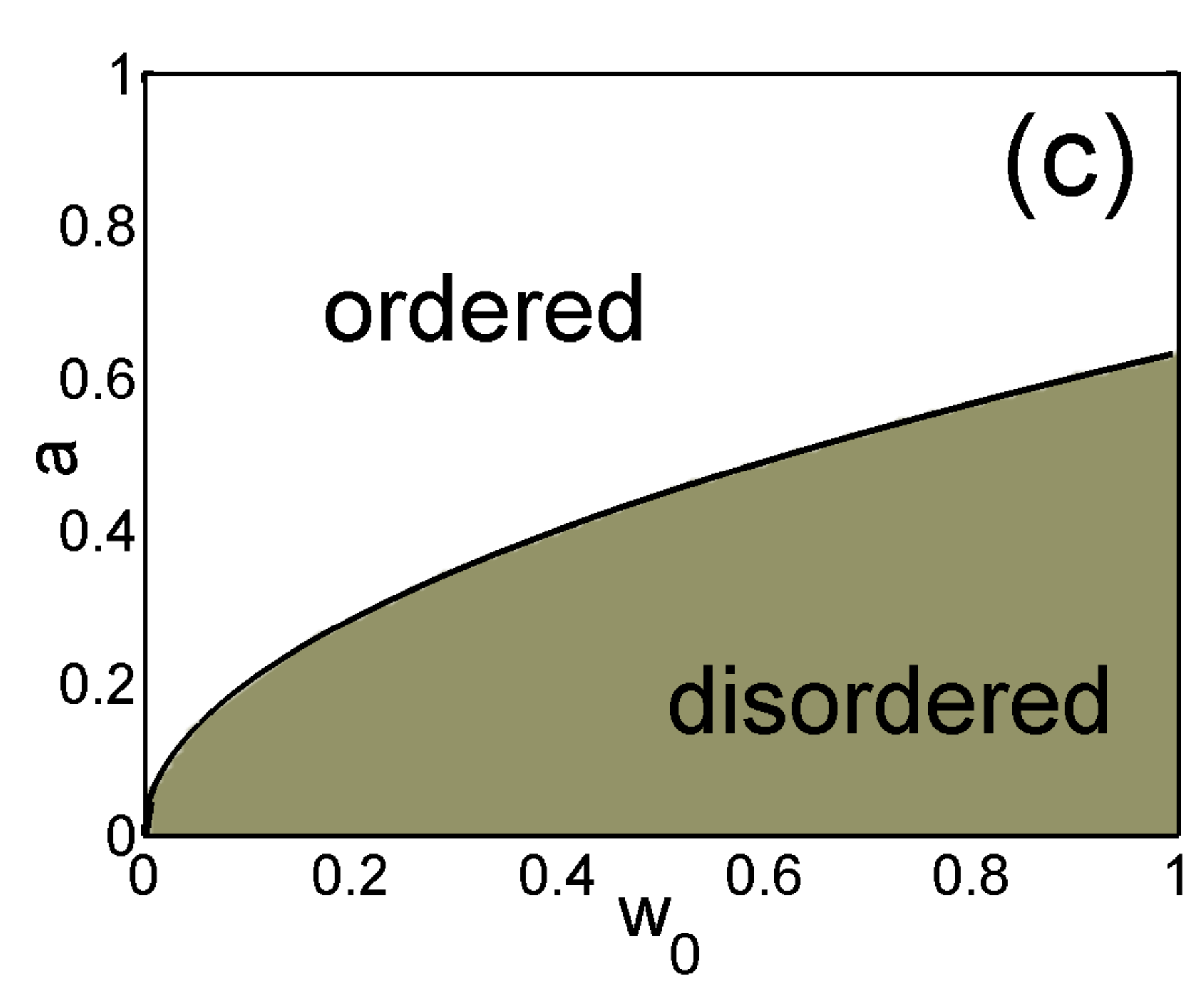}
\includegraphics[scale=0.225]{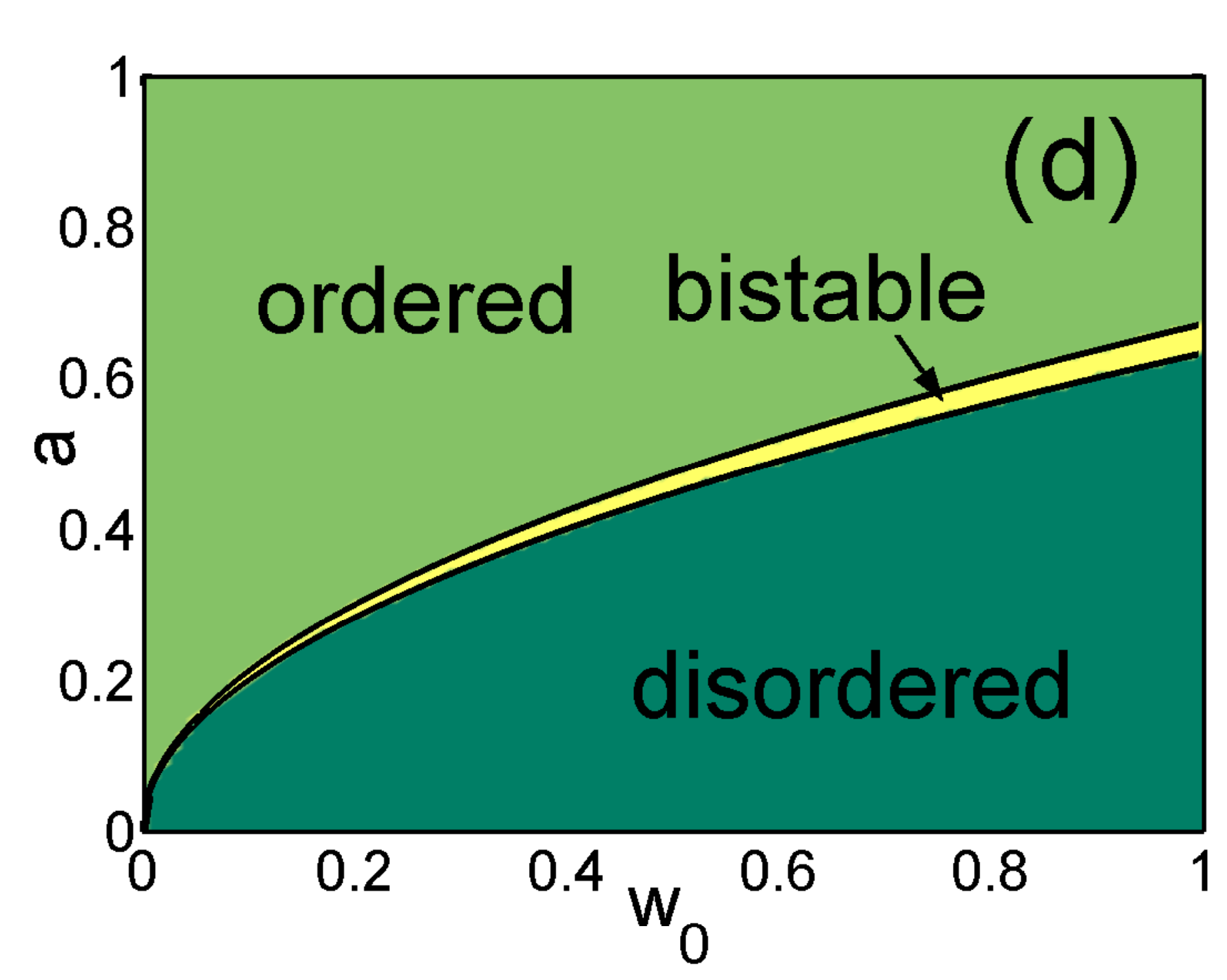}
\caption{(Color online). 
Bifurcation and phase diagrams of adaptive network systems with $M=2$ (left) and $M=3$ (right) available states per node. The bifurcation diagrams (top) show the density of nodes in a given state for the stable (solid) and unstable (dashed) stationary solutions. In both diagrams, the system undergoes a transition from a disordered solution to an ordered one as the noise level $w_0$ is decreased. 
For $M=2$ (a) this transition occurs through a supercritical pitchfork bifurcation and for $M=3$ (b), through a transcritical one, corresponding to a continuous or a discontinuous transition, respectively. Analytical results using a pair-level closure approximation (lines) are in good agreement with numerical network simulations (circles) using $N=10^4$ nodes. The phase diagrams (bottom) as a function of noise $w_0$ and link creation rate $a$ display a region of bistability only in the $M \geq 3$ case. Parameters: $a=0.5$ (top panels), $w_2=0.2$, $d=0.1$.}
\label{fig:3}
\end{figure}
%-----------------------------------------------------------------------------------------------------------------------------

To make analytical progress, we assume that the creation and deletion rates of every type of link cancel each other independently in the stationary solution, i.e.~$a [X][X']\!\!=\!\!d[XX']$. This is confirmed below by comparing our analytical results to direct agent-based numerical simulations of the full network dynamics. In analogy to the mean field case, we assume that all states have identical densities except for a single focal state. We denote by $[x]$ the density of nodes in this focal state and by $[j]$ the density of all other states.
Using this notation, we can rewrite Eq.(\ref{Eq:AN1}) as
\begin{equation}
\label{Eq:AN7}
\frac{d [x]\!}{d t}\!\!=\!\!w_0 \left( [j]\!-\![x] \right) \!+\!
            \frac{w_2}{2} \left( M\!-\!1 \right) [xj]^2 \left( \frac{1}{[j]} \!\!-\!\! \frac{1}{[x]} \right).
\end{equation}
By imposing the conservation law $\sum_{i=1}^M[i]=1$, we find the stationary solutions
\begin{equation}
\label{Eq:ANsol1}
[x]=[j]= 1 / M
\end{equation}
and
\begin{eqnarray}
\label{Eq:ANsol2a}
\left[x \right] &=& \frac{1 \pm \sqrt{1-w_0/c_1}}{2}, \\
\label{Eq:ANsol2b}
\left[j \right] &=& \frac{1 \mp \sqrt{1-w_0/c_1}}{2(M-1)}.
\end{eqnarray}
Here, $c_1= w_2 a^2 / (8 d^2)$ and Eq.~(\ref{Eq:ANsol2b}) represents $M-1$ identical equations for the node densities of all states other than $x$.

The results of the analysis (Fig.~\ref{fig:3}) are similar to those obtained with the mean field approximation: at low noise the disordered state becomes unstable and stable branches appear that correspond to the symmetry-broken solution. However, there are two differences.
First, only the ordered solution that has one majority opinion and $M-1$ minority opinions is stable; the reversed case with one minority opinion and $M-1$ majority opinions (set II in Fig.~\ref{fig:2}) is unstable.
Second, the bifurcation points now depend on the density of linked pairs, and are therefore a function of the link creation and deletion rates.
The saddle-node bifurcation (for $M > 2$) where the ordered states vanish now occurs at $c_1$, whereas the transcritical bifurcation where the disordered state loses its stability is at 
\begin{equation}
\label{Eq:c1c2}
c_2= \left[ 1\!-\!\left( \frac{M-2}{M} \right)^2 \right] c_1.
\end{equation}
%
% XXXXXXXXXXXXXXXXXXXXXXXXXXXXXXXXXXXXXXXXXXXXXXXXXXXXXXXXXXXXXXXXXXXXXXXXXXX
%
We therefore have $c_2 < c_1$ for all systems with more than two available states. Thus, the transition is generally of first order and has a bistable region in the $w_0$ interval given by $c_2 < w_0 < c_1$. 
In the limit of a large number of possible states, $c_2 \to 0$ and the region of bistability extends to the origin.
A continuous transition is only observed in the special case of two opinions, where $M=2$ implies $c_1=c_2$ and the two transitions coincide to form a pitchfork bifurcation.

These analytical predictions are in good agreement with results from large agent-based simulation runs (cf.~Fig.~\ref{fig:3} panels a,b). Only near the critical points a small difference is observed, which may be due to the moment closure approximation or finite size effects. 

%===============================================================
\section{Comparison to the Swarming Transition}
\label{Sec:Swarm}
%===============================================================

%------------------------------------------------------------------------------------------ Figure 4 ------------------------
\begin{figure}
\includegraphics[scale=0.3]{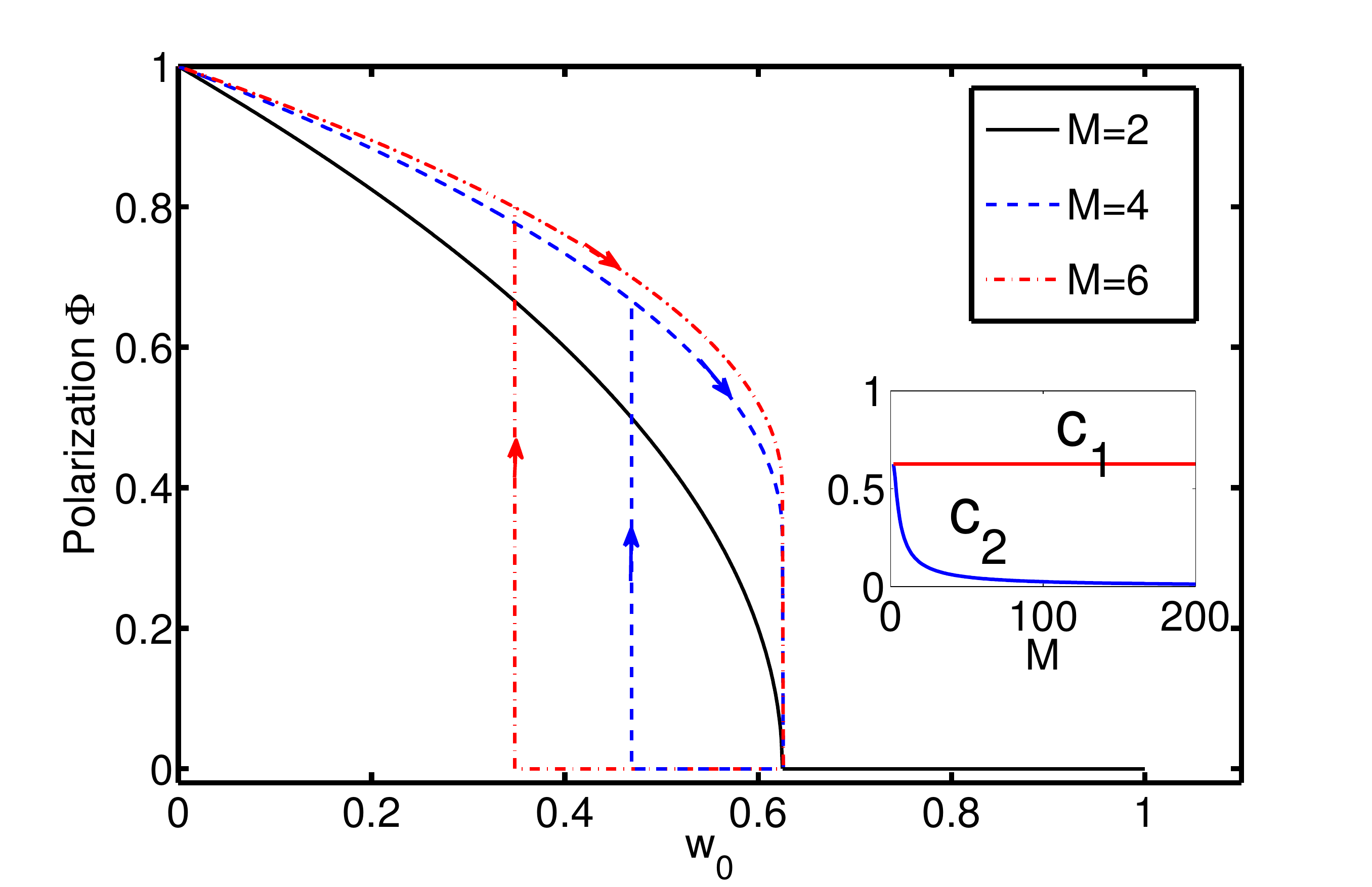}
\caption{
(Color online). Bifurcation diagrams of the order parameter $\Phi$ in Eq.~(\ref{Eq:psi2}) as a function of noise $w_0$.
The curves were computed using Eqs.~(\ref{Eq:ANsol1}-\ref{Eq:ANsol2b}) for $M = 2$, $4$, and $6$ potential heading directions, corresponding to $D=1$, $2$, and $3$ dimensions, respectively. 
For $M = 2$, the transition is continuous. For $M > 2$, the critical value of the control parameter where the $\Phi = 0$ branch loses stability ($c_2$) becomes smaller than the point where the upper branch vanishes ($c_1$). This results in a discontinuous transition and a region of bistability, which gives rise to the hysteresis cycles indicated by the arrows. 
The inset displays $c_1$ and $c_2$ as a function of $M$; 
the bistable region is broader for larger $M$ values.
}
\label{fig:4}
\end{figure}
%-----------------------------------------------------------------------------------------------------------------------------

In this Section, we will relate the ordering transition described above to the collective motion transition observed in swarms \cite{VicsekPR}.
For this purpose, each node is interpreted as a self-propelled agent, its state as its heading direction, and linked nodes as interacting agents. An advantage of this approach is that it does not require specifying the details of the interactions. While collective motion can result from a broad variety of interactions (such as aligning \cite{Vicsek1}, attraction-repulsion \cite{Ferrante1,Ferrante2}, or escape-pursuit \cite{Pawel}), the adaptive-network perspective can treat all of these equally by focusing on the exchanges of information that lead to consensus on the collective heading direction, without considering the details of the interactions.
In particular, if we assume that interactions can only occur within a given distance, the limit case studied in this paper (where the linking and unlinking rates between agents in the same state is set to zero) can be mapped to a situation where agents that advance in a common direction do not change their relative positions and therefore do not create or destroy interactions between them.
We also focus on the simplest limit case where the linking and unlinking rates (corresponding here to the encounter and disbanding rates at which agents start or stop interacting with each other) are constant and equal for all agents in different states (i.e. with different headings).

In order to compare our adaptive network system to collective motion, we start by associating each node state $[h]$ with an agent's heading $\hat{v}$ in a space where agents can only move in discretized directions that are perpendicular to each other. Each $\hat{v}$ is thus a unit vector pointing in a direction that is either opposite or orthogonal to all others. The number of potential headings $M$ therefore depends on the dimensionality of the space, with $M=2$ in one dimension, $M=4$ in two dimensions, $M=6$ in three dimensions and, in general, $M=2 D$ in $D$ dimensions.
The usual polarization order parameter used to describe the degree of alignment and of collective motion in swarming systems is given by
\begin{equation}
\label{Def:psi}
\Phi = \frac{1}{N} \left| \sum_{i=1}^N \hat{v}_i \right|,
\end{equation}
where $N$ is the total number of agents and $\hat{v}_i$ is a unit vector indicating the heading direction of agent $i$ \cite{VicsekPR}.
With this definition, $\Phi = 1$ if all agents are perfectly aligned and swarming in the same direction, whereas $\Phi = 0$ if they are randomly oriented.
In the discretized space with only orthogonal heading directions that we consider here, $\Phi$ can be expressed as
\begin{equation}
\label{Eq:psi2}
\Phi = \sqrt{ \sum_{h=1}^{M/2} \left( \left[ 2 h \right] - \left[ 2 h - 1 \right] \right)^2 }.
\end{equation}
This relationship allows us to plot the polarization order parameter $\Phi$ as a function of 
$w_0$, which serves as a proxy for the amount of noise in the agent motion (Fig.~\ref{fig:4}).

In the context of swarms, the symmetry-breaking bifurcations computed above correspond to ordering phase transitions to collective motion. For agents moving in one-dimensional space ($M=2$), the transition is continuous (second order), whereas for agents moving in more dimensions ($M \geq 4$), the transition is discontinuous (first order). A region of bistability appears for $M \geq 3$ and becomes broader for higher values of $M$ (Fig.~\ref{fig:4}, inset).  

Although the adaptive network approach includes several approximations, the results above  provide insights into the more complex problem of understanding general features of the transition to collective motion in swarms.
The question of whether the actual swarming transition to collective motion is continuous or discontinuous, for example, has been the subject of intense debate \cite{Vicsek1}.
While in the initial numerical explorations the transition appeared to be continuous (second order), it was later shown through theoretical arguments and large-scale numerical simulations that it is, in fact, discontinuous (first order) and has a bistable transition region where ordered and disordered swarming states coexist \cite{Vicsek1,Chate1, Chate2}.
The results presented in Fig.~\ref{fig:4} would suggest that, generically, this transition should be continuous in one dimension and discontinuous in two or three dimensions, with a more prominent bistable region in the 3d case.

To the best of our knowledge, there has been no systematic analysis of the properties of the ordering transition as a function of the embedding space dimensionality for different types of swarming models. In one dimension, various approaches have concluded that the transition is either absent or first order \cite{1DCollMotA,1DCollMotB,1DCollMotC,1DCollMotD}. In two dimensions, the transition has been much better studied and shown to be first order, as in three dimensions, but the size of their bistable regions has not been compared \cite{VicsekPR,TT2}.

%------------------------------------------------------------------------------------------ Figure 5 ------------------------
\begin{figure}[tp]
\includegraphics[scale=0.225]{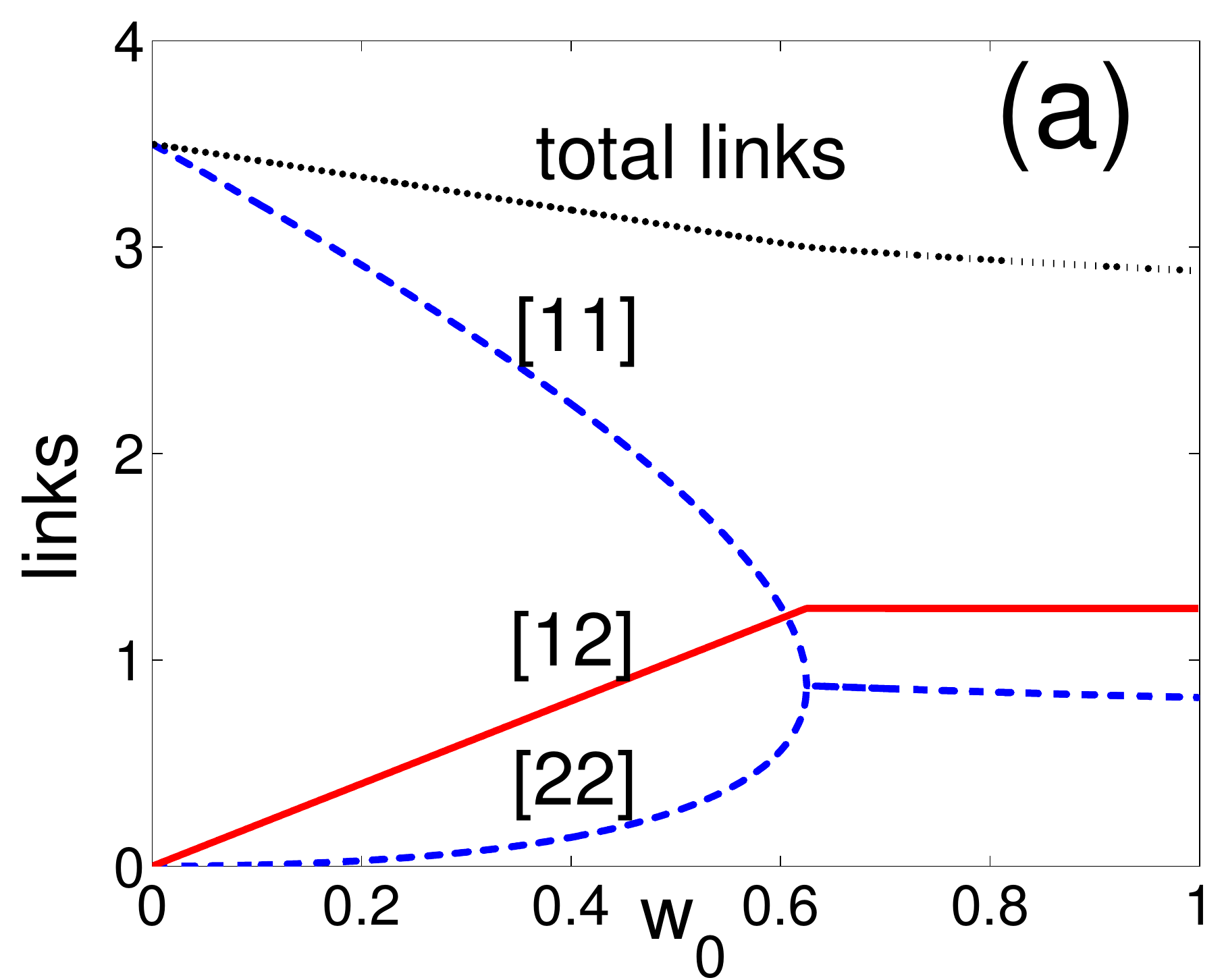}
\includegraphics[scale=0.225]{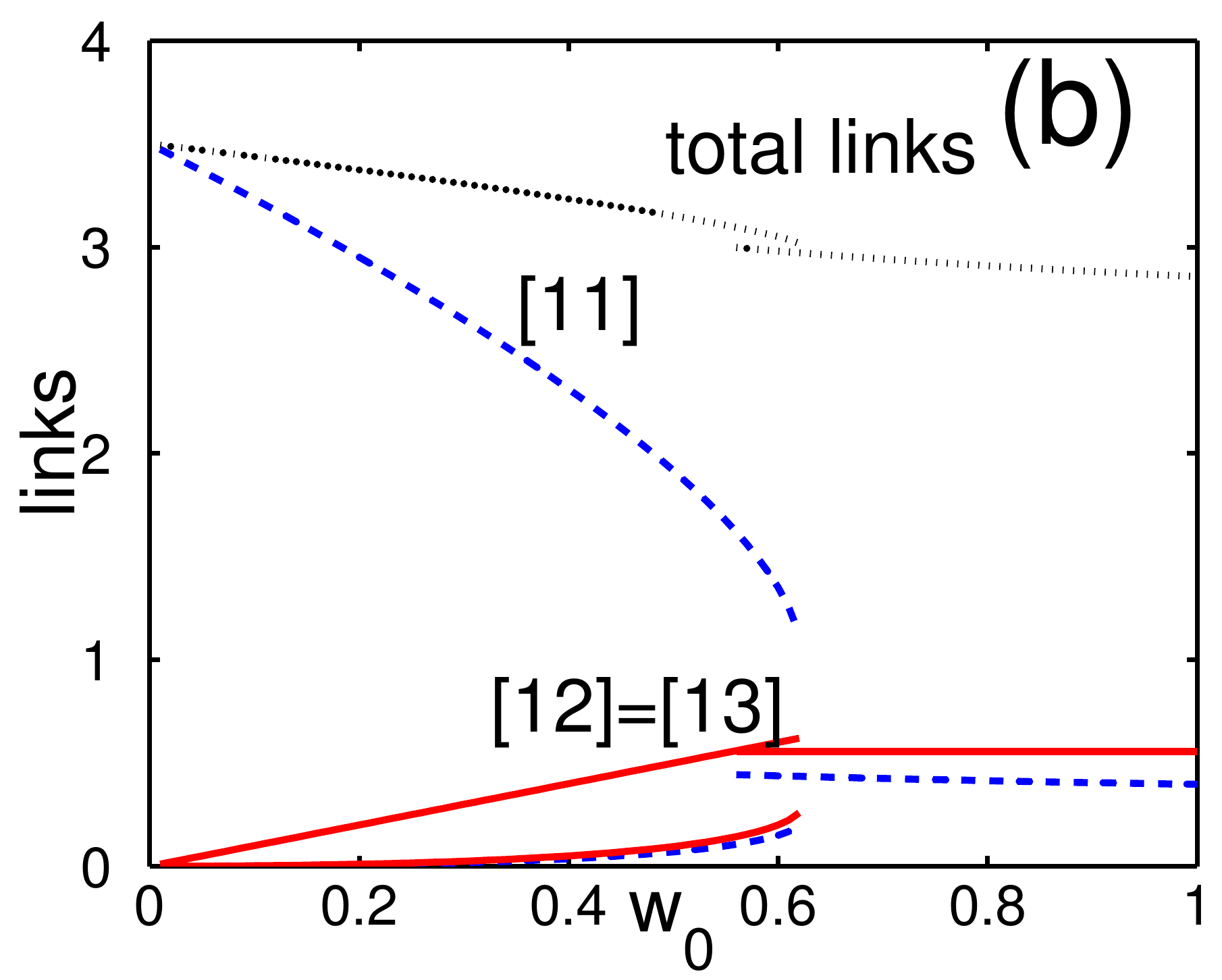}
\caption{(color online). Per-capita density of linked pairs as a function of noise $w_0$, computed analytically for $M=2$ (a) and 3 (b) using the same parameters as in the corresponding bifurcation diagrams in Fig.~\ref{fig:3}.
Blue dashed lines: Density of linked pairs with both nodes in the same state of majority (labeled $[11]$) or minority opinion ($[22]$ and $[33]$, the lowest branch at left side of the bifurcation in both plots). 
Red solid lines: Density of linked pairs with one node in the majority and one in the minority opinion ($[12]$ and $[13]$), or both in the minority opinions ($[23]$ in panel (b), the lowest red line at the left side of the bifurcation).
Black dotted lines: Total density of linked pairs.
The bifurcation features displayed in Fig.~\ref{fig:3} are mirrored here in these link density plots.}
\label{fig:5}
\end{figure}
%-----------------------------------------------------------------------------------------------------------------------------

We can further examine the connection between our adaptive network model and swarming systems by considering the per-capita densities of linked pairs displayed in Fig.~\ref{fig:5}. 
These match the interaction frequencies that are expected to occur in swarms. 
For example, the total number of links decreases monotonically with noise level, which corresponds to the observation that higher noise values will produce less clustering and therefore fewer interactions between agents in swarming systems \cite{HuepeAldanaPRL,VicsekPR}.
We also see that the density of heterophilic links $[12]$ (and $[13]$ in the $M\!=\!3$ case) increases with noise. This can be explained by an increasing rate of encounters at higher noise levels.

Furthermore, we find that for all cases with $M>2$ (such as the $M=3$ case displayed in the figure) the density of heterophilic links in the ordered branch, $[12]$ and $[13]$, continues to increase as a function of noise within the bistable region, where it becomes higher than that of the disordered branch. 
Despite this high number of heterophilic links, the ordered branch persists because the density of homophilic links $[11]$ is also high.
This can be related to what is observed in the bistable region in swarms, where it is known that a higher density of interactions between agents in the majority heading state, which corresponds to the formation of high-density bands (oriented perpendicular to the heading direction) in two or more dimensions, stabilizes the ordered state \cite{Chate1, Chate2}, leading to a bistable region and thus to a discontinuous transition.
This analogy could provide an alternative way to understand the details of the swarming transition as a function of the dimension of the embedding space. 
%

%===============================================================
\section{Conclusions}
\label{Sec:Conclusions}
%===============================================================

In this paper, we analyzed the swarming systems class of adaptive network models, where links can only be created or deleted between nodes in different states.
We showed analytically that this class displays a symmetry-breaking transition with properties that depend on the number of states $M$ accessible to each node.
If $M=2$, the transition occurs through a supercritical pitchfork bifurcation; if $M \geq 3$, through a subcritical one. Consequently, only this latter case displays a bistable region near the bifurcation point.
Note, however, that previous work \cite{Cristian} had shown that bistable solutions can also be obtained in the $M=2$ case if we allow link creation and deletion processes to occur between nodes in the same state, a situation that was not studied here.

The results above, taken together, provide insights on a potential direct connection between link dynamics, their dependence on internal states, and the resulting properties of this type of symmetry-breaking transitions.
We also discussed in this paper their implications for the analysis of the collective motion transition in swarms.

The parallels between the adaptive-network approach presented here and agent-based dynamics are not restricted to swarming systems. They can be extended to any group of agents moving in an abstract phase space with similar dynamical rules. These rules must consider agents with an internal state (as the heading direction in the swarming case) that determines their trajectory in this phase space, in which their relative positions determine whether they interact. 
It is thus conceivable that the proposed model could be extended to study social processes involving heterophily, such as the diffusion of innovations and technologies \cite{Rogers} or job seeking through weak interpersonal ties \cite{Granovetter}.

We would like to thank Gerd Zschaler and G\"uven Demirel for their assistance in using the XPPAUT and largetnet library.
The work of CH was supported by the US National Science Foundation under Grant No. PHY-0848755. 
No empirical data was produced in this work. Additional material will be made available on www.biond.org.


\begin{thebibliography}{}
\bibitem{footnote1}
It is also natural to consider an additional process, whereby a pair of linked nodes $X-Y$, each in a different state, can influence each other to produce either $X-X$ or $Y-Y$. For the stationary solutions considered here, however, these terms will cancel out and the total density of nodes in a given state will remain unchanged \cite{Cristian}. For simplicity, we will therefore not include here any terms representing these processes.
\end{thebibliography}
\end{document}